\documentclass[11pt]{article}
\usepackage{axodraw}
\usepackage{graphicx}
\usepackage{amssymb}
\usepackage{epstopdf}
\def\ch{{\rm ch}}
\def\sh{{\rm sh}}
\def\ba{\begin{eqnarray}}
\def\ea{\end{eqnarray}}
\def\nn{\nonumber}

\def\tlscbf{\begin{picture}(12,12)(12,12)
\Photon(4,15)(26,15)2 5
\CArc(15,15)(10,0,180)
\CArc(15,15)(10,180,360)
\CArc(15,15)(12,0,180)
\CArc(15,15)(12,180,360)
\end{picture}}

\def\olmass{\begin{picture}(12,12)(12,12)
\PhotonArc(15,10)(10,0,180)2 5
\Line(0,10)(30,10)
\end{picture}}

\def\tlscbfph{\begin{picture}(12,12)(12,12)
\DashLine(4,15)(26,15){1}
\CArc(15,15)(10,0,180)
\CArc(15,15)(10,180,360)
\CArc(15,15)(12,0,180)
\CArc(15,15)(12,180,360)
\end{picture}}

\def\tlscbfphv{\begin{picture}(12,12)(12,12)
\DashLine(4,15)(26,15){1}
\CArc(15,15)(10,0,180)
\CArc(15,15)(10,180,360)
\CArc(15,15)(12,0,180)
\CArc(15,15)(12,180,360)
\Vertex(15,15)3
\end{picture}}

\def\tlspbf{\begin{picture}(12,12)(12,12)
\Photon(4,15)(26,15)2 5
\CArc(15,15)(10,0,180)
\CArc(15,15)(10,180,360)
\CArc(15,15)(11,0,180)
\CArc(15,15)(11,180,360)
\CArc(15,15)(12,0,180)
\CArc(15,15)(12,180,360)
\end{picture}}

\def\tlscfr{\begin{picture}(12,12)(12,12)
\Photon(5,15)(25,15)2 5
\CArc(15,15)(10,0,180)
\CArc(15,15)(10,180,360)
\end{picture}}

\def\tlscfrph{\begin{picture}(12,12)(12,12)
\DashLine(5,15)(25,15){1}
\CArc(15,15)(10,0,180)
\CArc(15,15)(10,180,360)
\end{picture}}

\def\olscbf{\begin{picture}(12,12)(12,12)
\CArc(15,15)(10,0,180)
\CArc(15,15)(10,180,360)
\CArc(15,15)(12,0,180)
\CArc(15,15)(12,180,360)
\end{picture}}

\def\olscfr{\begin{picture}(12,12)(12,12)
\CArc(15,15)(10,0,180)
\CArc(15,15)(10,180,360)
\end{picture}}

\def\olscbfv{\begin{picture}(12,12)(12,12)
\CArc(15,15)(10,0,180)
\CArc(15,15)(10,180,360)
\CArc(15,15)(12,0,180)
\CArc(15,15)(12,180,360)
\Vertex(15,4)3
\end{picture}}

\def\olscfrv{\begin{picture}(12,12)(12,12)
\CArc(15,15)(10,0,180)
\CArc(15,15)(10,180,360)
\Vertex(15,5)3
\end{picture}}

\title{Two-Loop Diagrammatics in a Self-Dual Background}
\author{Gerald V. Dunne\\
Department of Physics\\
University of Connecticut\\
Storrs, CT 06269-3046 USA}
\begin{document}
\maketitle

\begin{abstract}
Diagrammatic rules are developed for simplifying two-loop QED diagrams with propagators in a constant self-dual background field.
This diagrammatic analysis, using dimensional regularization, is used to explain how the fully renormalized two-loop Euler-Heisenberg effective Lagrangian for QED in a self-dual background field is naturally expressed in terms of one-loop diagrams. The connection between the two-loop and one-loop  vacuum diagrams in a  background field parallels a corresponding connection for free vacuum diagrams, without a background field, which can be derived by simple algebraic manipulations. It also mirrors similar behavior recently found for two-loop amplitudes in N=4 SUSY Yang-Mills theory.
\end{abstract}

\section{Introduction}

There has been dramatic progress in recent years in various aspects of higher-loop quantum field theory. Many two-loop scattering amplitudes in QED, QCD and SUSY theories are now computable \cite{steinhauserreview,gloverreview,bernreview}. This recent progress has been made possible by techniques developed over many years, such as: integration-by-parts \cite{vladimirov,chetyrkin,broadhurst}, helicity and color decompositions \cite{mangano,bkcolor,dixontasi}, world-line methods \cite{bernkosower,schubertzakopane,chrisreview}, collinear and infrared factorizations \cite{collinear,catani,sterman}, computerized algabraic manipulations \cite{vermaserenform}, and differential \cite{kotikov,remiddi},  difference \cite{laporta} and recurrence \cite{baikov,smirnovsteinhauser} relations for diagrams. Analytic expressions for "master diagrams" \cite{remiddireview}, such as the two-loop ones in \cite{smirnov,tausk}, have played a crucial role in obtaining one-loop and two-loop amplitudes for scattering processes in gauge theories such as QED \cite{qed} and QCD \cite{qcd}, as well as in gravity \cite{berngravity} and SUSY \cite{susy} theories. Furthermore, the developments in computer algebra have enabled the 4-loop computations of the $\beta$-function and anomalous mass dimension in QED \cite{gorishnii} and QCD \cite{4loop}. For vacuum diagrams, the reduction to, and  evaluation of, master diagrams has been extended to the four-loop level in nonabelian gauge-Higgs systems \cite{schroder}.
In addition to these impressive computational advances, there is also now a much better understanding of the algebraic properties of Feynman diagram computations, due to breakthroughs of Kreimer et al \cite{dirk}, who have discovered a powerful Hopf algebra structure underlying Feynman perturbation theory. 
There has also been significant recent progress in computing higher loop effective actions. This has been driven in part by advances in the string-inspired world-line approach \cite{bernkosower,schubertzakopane,chrisreview}, which has been extended beyond one-loop \cite{strassler} to the multi-loop level \cite{Sch1}. Euler-Heisenberg effective actions at two-loop have now been studied in QED \cite{ritus,dittrich,mss,holger,dunsch,ds1,ds2,ds3,dgs}, scalar theories \cite{satophi3}, QCD \cite{satoym} and supersymmetric theories \cite{kuzenko}. Since the effective action is the generating function for scattering amplitudes, these developments for two-loop amplitudes and two-loop effective actions calculations are closely related. 

The main goal of this paper is to begin to develop diagrammatic rules for the manipulation of vacuum diagrams involving propagators {\it in a background field}, along the lines of the now familiar techniques for the manipulation of diagrams involving {\it free} propagators. 
This should facilitate the analysis of higher-loop background field vacuum diagram calculations. These background field vacuum diagrams arise in the computation of effective actions, and can be used as generating functions for computing multi-leg amplitudes \cite{ds2,louise}. More specifically, we consider two-loop QED vacuum diagrams in a constant self-dual background. There are several hints that simple diagrammatic rules may be attainable in this case.
First, it has recently been shown that the two-loop QED Euler-Heisenberg effective action in a self-dual background is extremely simple, after mass and charge renormalization, and is found to be expressed in terms of simpler one-loop diagrams \cite{ds1,ds2,dgs}.  This paper gives a diagrammatic analysis of this result, using some of the ideas developed for the manipulation of diagrams involving free propagators. We show that one of the coefficients is found to reflect a similar two-loop to one-loop relation in the free theory (with no background field), which can be derived by trivial algebraic manipulations without any integrals. Second, the effective action in a self-dual background generates amplitudes with all external helicities alike, and these amplitudes are known to be especially simple in massless theories \cite{dixontasi,bardeen}. While less is known about helicity amplitudes with massive fermions or scalars, there are still simplifications \cite{bernmorgan}.
Third, self-duality is closely tied to both helicity and SUSY \cite{dgs}, and amplitudes in SUSY theories are known to be simple \cite{susy}. Indeed, recent studies of two-loop amplitudes in N=4 SUSY Yang-Mills theory have found remarkable relations between the two-loop results and their one-loop counterparts \cite{abdk}. It is anticipated that this is a perturbative reflection of the nonperturbative solvability of these highly symmetric theories (an illustration of the Maldacena conjecture \cite{maldacena}), and that similar recursive structures will continue to higher loop order \cite{abdk}. This should be reflected in the effective action, perhaps by a mechanism similar to the one described in this paper. If there is some recursive or algebraic structure in this background field perturbation theory, it opens a window of opportunity to learn something nonperturbative about gauge theories such as QED and QCD.

Diagrammatic methods in background field calculations have also been investigated recently \cite{bornsen} in a computation of the three loop Yang-Mills $\beta$-function. There, since the goal is the $\beta$-function rather than the full effective action, it is sufficient to expand the background field propagators to at most quadratic order in the background field strength \cite{shifman}. Thus, in this approach one actually deals with diagrams involving powers of the free propagators, in which case the conventional diagrammatic rules can be used for evaluating diagrams. Here, instead, we keep the full dependence on the background field, in order to obtain the full effective action, the generating function of amplitudes with any number of external legs. This requires the development of new diagrammatic rules. 

The computation \cite{he,schwinger} of the one-loop  Euler-Heisenberg effective Lagrangian for QED in a constant background field serves as the paradigm for low energy effective field theory, with many immediate applications, such as nonlinear effective interactions ({\it e.g.}, light-light scattering), pair production from vacuum in an electric background, and the $\beta$-function. 
In principle, the computation of the two-loop Euler-Heisenberg effective Lagrangian is completely straightforward, as we only need to compute a single vacuum diagram with an internal photon line and a single fermion or scalar loop, where these fermion/scalar propagators are those in the presence of the constant background field. These propagators have been known in closed-form for a long time \cite{fock,schwinger}. The new feature that makes two-loops nontrivial is the necessity of mass renormalization. In a {\it tour de force}, Ritus \cite{ritus} found explicit expressions for the renormalized two-loop effective Lagrangian for both spinor and scalar QED. These two-loop results are now well understood, having been studied using proper-time regularization, dimensional regularization and more recently using the world-line formalism \cite{dittrich,mss}. Unfortunately, the two-loop effective Lagrangian expressions, for a general constant background, are extremely complicated two-parameter integrals (roughly speaking, there is one parameter integral left for each spinor/scalar propagator), and it is difficult to extract detailed physical information beyond their leading behavior. This complication persists even if one specializes to the less general case of a constant magnetic field or a constant electric field background \cite{ritus,dittrich,dunsch}.

However, if the constant background field is chosen to be self-dual, then at two loops all the parameter integrals can be done in closed form, leading to the following
explicit formulas for the fully renormalized two-loop effective Lagrangians \cite{ds1,ds2}:
\ba
{\cal L}_{\rm scalar}^{(2)}
=
\alpha \,{m^4\over (4\pi)^3}\frac{1}{\kappa^2}\left[
{3\over 2}\xi^2 (\kappa)
-\xi'(\kappa)\right]
\label{2lsc}
\ea
\ba
{\cal L}_{\rm spinor}^{(2)}
=
-2\alpha \,{m^4\over (4\pi)^3}\frac{1}{\kappa^2}\Bigl[
3\xi^2 (\kappa)
-\xi'(\kappa)\Bigr]
\label{2lsp}
\ea
These on-shell renormalized two-loop effective Lagrangians are expressed in terms of the dimensionless variable $\kappa=\frac{m^2}{2ef}$, where $f$ denotes the self-dual field strength: $\frac{1}{4}F_{\mu\nu}F_{\mu\nu}=f^2$. The function $\xi$ is defined as
\ba
\xi(\kappa)\equiv -\kappa\Bigl(\psi(\kappa)-\ln(\kappa)+{1\over 2\kappa}\Bigr)
\label{xi}
\ea
where $\psi(\kappa) = {d\over d\kappa}{\rm ln}\,\Gamma(\kappa)$
is the Euler digamma function \cite{bateman}. The subtraction of the first two terms of the asymptotic expansion of $\psi(\kappa)$ in the definition of $\xi(\kappa)$ arises naturally in the renormalization, as is shown below.

The dramatic simplicity of the results (\ref{2lsc},\ref{2lsp}) for a self-dual background field, compared to the complicated double-parameter integral representations obtained for a general background field \cite{ritus,dittrich,mss}, can be traced to the relationship between self-duality, helicity and supersymmetry \cite{ds2,ds3,dgs}. Aiming to go beyond the two-loop level, this paper makes connection of these results with the diagrammatic methods developed, using  dimensional regularization, for higher loop amplitudes in quantum field theory. We restrict attention here to the case of scalar QED, leaving some interesting issues concerning the zero modes of spinor QED in a self-dual background to a future publication. 

\section{Scalar Propagator in a Self-Dual Background}
The propagators of scalar and spinor particles in a constant background electromagnetic field, of field strength $F_{\mu\nu}$, may be expressed as proper-time integrals \cite{fock,schwinger}. If the electromagnetic background is self-dual, so that (we work in Euclidean space)
\begin{eqnarray}
F_{\mu\nu}=\tilde{F}_{\mu\nu}\quad , \quad {\rm where} \quad \tilde{F}_{\mu\nu}\equiv \frac{1}{2}\epsilon_{\mu\nu\rho\sigma}F_{\rho\sigma}
\label{selfdual}
\end{eqnarray}
then the constant field-strength matrix $F_{\mu\nu}$ may be block-diagonalized as
\ba
F_{\mu\nu}=\left(\matrix{0&f&0&0\cr -f&0&0&0\cr 0&0&0&f\cr 0&0&-f&0}\right)
\label{block}
\ea
where $f$ denotes the strength of the constant field: $\frac{1}{4}F_{\mu\nu}F_{\mu\nu}=f^2$. The block-diagonalized form of $F_{\mu\nu}$ also implies that $F_{\mu\nu}F_{\nu\rho}=-f^2\delta_{\mu\rho}$, and this fact leads to a dramatic simplification of the expressions for the scalar and spinor propagators in a constant electromagnetic background. For example, in 4 dimensions the scalar propagator has the proper-time integral representation
\begin{eqnarray}
G(p)=\int_0^\infty \frac{dt}{\ch^{2}(eft)} \, e^{-m^2 t-\frac{p^2}{ef}\tanh(eft)}
\label{4dscprop}
\end{eqnarray}
The massless propagator, $(1-e^{-p^2/ef})/p^2$, has been studied in one-loop calculations in \cite{leutwyler}. 

The self-duality condition (\ref{selfdual}) is naturally tied to 4 dimensional spacetime. However, the block-diagonalization in (\ref{block}) generalizes straightforwardly to any integer dimension, retaining the property $F_{\mu\nu}F_{\nu\rho}=-f^2\delta_{\mu\rho}$ if the dimension is even. Thus, the natural dimensional continuation of the scalar propagator  for the case of a block-diagonalized "self-dual" constant background field $F_{\mu\nu}$ is
\begin{eqnarray}
G(p)=\int_0^\infty \frac{dt}{\ch^{d/2}(eft)} \, e^{-m^2 t-\frac{p^2}{ef}\tanh(eft)}
\label{scprop}
\end{eqnarray}
This corresponds to standard dimensional regularization within the world-line formalism \cite{schubertzakopane,chrisreview,mss}. 

The propagator $G(p)$ in (\ref{scprop}) satisfies the following differential equation, for arbitrary dimension $d$:
\ba
\left(p^2+m^2\right)G(p)=1+\frac{(ef)^2}{4} \frac{\partial^2 G(p)} {\partial p^2_\mu} 
\label{propde}
\ea
In the free-field limit, where $f\to 0$, the propagator $G(p)$ in (\ref{scprop}) reduces smoothly to the free propagator 
\ba
G_0(p)=\frac{1}{p^2+m^2}
\label{freeprop}
\ea
as is clearly consistent with the differential equation (\ref{propde}).


An important observation is that the propagator $G(p)$ in (\ref{scprop}) is naturally connected with the function $\xi(\kappa)$ defined in (\ref{xi}), in terms of which the two-loop renormalized effective Lagrangians (\ref{2lsc},\ref{2lsp}) are expressed. To see this, using an integral representation \cite{bateman} of the digamma function $\psi(\kappa)$, we can write $\xi(\kappa)$ as:
\ba
\xi(\kappa)=- \frac{1}{2}\int_0^\infty ds\, e^{-2\kappa s}\left(\frac{1}{\sh^2 s}-\frac{1}{s^2}\right)
\label{xirep}
\ea
Next, introducing the notation that a double line represents the scalar propagator in the self-dual background, while a single line represents the free scalar propagator, it is simple to show that 
\ba
\mbox{\olscbf}\,\,\, -\quad\mbox{\olscfr} \,\,\, &\equiv &\int \frac{d^4p}{(2\pi)^4}\left[G(p)-G_0(p)\right]=-\frac{m^2}{(4\pi)^2} \,\frac{ \xi(\kappa)}{\kappa}
\label{xiint}
\ea
Further, introducing the notation that a solid dot on a propagator means that the propagator is squared, we find that
\ba
\mbox{\olscbfv}\,\,\, -\quad\mbox{\olscfrv} \,\,\, &\equiv &\int \frac{d^4p}{(2\pi)^4}\left[(G(p))^2-(G_0(p))^2\right]= \frac{\xi^\prime(\kappa)}{(4\pi)^2}
\label{xiprint}
\ea
Thus, we can express the results (\ref{2lsc},\ref{2lsp}) for the renormalized two-loop effective Lagrangians in scalar and spinor QED diagrammatically as
\ba
{\rm scalar\,\,QED:}\,\, \Bigl[\quad\mbox{\tlscbf}\,\,\, - \quad \mbox{\tlscfr}\,\,\Bigr]&=&\frac{3}{2}\,e^2\, \Bigl[\quad \mbox{\olscbf}\,\,\, - \quad \mbox{\olscfr}\,\,\, \Bigr]^2 -e^2\frac{(ef)^2}{4\pi^2}  \Bigl[\quad \mbox{\olscbfv}\,\,\, -\quad  \mbox{\olscfrv}\,\,\, \Bigr]
\label{2lscdig}
\ea
\ba
{\rm spinor\,\,QED:} \,\, \Bigl[\quad\mbox{\tlspbf}\quad - \quad \mbox{\tlscfr}\,\,\, \Bigr]&=&-6\,e^2\, \Bigl[\quad \mbox{\olscbf}\,\,\, - \quad \mbox{\olscfr}\,\,\,\Bigr]^2 +e^2 \frac{(ef)^2}{2\pi^2} \Bigl[\quad \mbox{\olscbfv}\quad -\quad  \mbox{\olscfrv}\,\,\, \Bigr]
\label{2lspdig}
\ea
In (\ref{2lspdig}) the triple line denotes the spinor propagator in the self-dual background. 
Note that all diagrams on the right-hand side are one-loop diagrams, which are much simpler than the original two-loop diagrams on the left-hand side.
The results (\ref{2lsc},\ref{2lsp}) have been derived previously using the world-line formalism \cite{ds1,ds2}, and in field theory with a proper-time cutoff \cite{dgs}. The purpose of this paper is to understand how these results come about in a diagrammatic approach, using conventional dimensional regularization.

\section{Bare Two Loop Effective Lagrangians}

Consider the two-loop vacuum diagram in scalar QED, with free scalar propagators and using Feynman gauge for the internal photon line. Simple algebraic manipulations reduce this two-loop diagram as follows:
\ba 
\tlscfr \,\, &=&\frac{e^2}{2} \int \frac{d^dp\, d^dq}{(2\pi)^{2d}}\, \frac{(p+q)^2}{(p-q)^2(p^2+m^2)(q^2+m^2)}\nn \\
&=& \frac{e^2}{2} \int \frac{d^dp\, d^dq}{(2\pi)^{2d}}\, \frac{\left[-(p-q)^2+2(p^2+m^2)+2(q^2+m^2)-4 m^2\right]}{(p-q)^2(p^2+m^2)(q^2+m^2)}\nn \\
&=& -\frac{e^2}{2}\left[\int \frac{d^dp}{(2\pi)^d} G_0(p)\right]^2 -2 e^2 m^2  \int \frac{d^dp\, d^dq}{(2\pi)^{2d}}\, \frac{G_0(p) G_0(q)}{(p-q)^2}\nn \\
&=& -\frac{e^2}{2}\Bigl[\quad \olscfr \,\,\, \Bigr]^2 -2 e^2 m^2  \Bigl[ \quad \tlscfrph \,\,\, \Bigr]
\label{freemanip}
\ea
In going from the second to third line we have used the fact that in dimensional regularization $\int d^dp\, (p^2)^\alpha =0$.

In fact,  the second term in the last line of (\ref{freemanip}) is proportional to the first term. This  follows most simply by an integration by parts \cite{vladimirov,chetyrkin,broadhurst}, but can also be seen by direct evaluation. In $d=4-2\epsilon$ dimensions it is straightforward to show that
\ba
\olscfr\,\,\, = \int \frac{d^dp}{(2\pi)^d}\, G_0(p)= -\frac{(m^2)^{1-\epsilon}}{(4\pi)^{2-\epsilon}}\, \frac{\Gamma(1+\epsilon)}{\epsilon(1-\epsilon)}
\label{oneloopfreeprop}
\ea
while
\ba
\tlscfrph\,\,\, =
\int \frac{d^dp\, d^dq}{(2\pi)^{2d}}\, \frac{G_0(p) G_0(q)}{(p-q)^2}&=&
-\frac{(m^2)^{1-2\epsilon}}{(4\pi)^{4-2\epsilon}} \frac{(\Gamma(1+\epsilon))^2}{\epsilon^2(1-\epsilon)(1-2\epsilon)}\nn\\
&=& -\frac{1}{m^2}\frac{(1-\epsilon)}{(1-2\epsilon)} \Bigl[\quad \olscfr\,\,\, \Bigr]^2
\label{tlfr}
\ea
Inserting these results into (\ref{freemanip}) we find
\ba 
\tlscfr\,\,\,  = \frac{e^2}{2} \left(\frac{d-1}{d-3}\right) \Bigl[\quad \olscfr\,\,\,  \Bigr]^2
\label{freemanipfull}
\ea
Thus, the two-loop vacuum diagram has been expressed as the square of a simpler one-loop vacuum diagram.
Notice that the dimension-dependent numerical prefactor, which gives a factor of $\frac{3}{2}$ in $d=4$, corresponds precisely to the factor of $\frac{3}{2}$ in the self-dual two-loop effective Lagrangian (\ref{2lscdig}). A similar argument also works for the factor of $-6$ in (\ref{2lspdig}) for the spinor case.

Now repeat the same type of manipulation as in (\ref{freemanip}), but for the two-loop vacuum diagram {\it in a self-dual background  field}. The scalar propagators change from $G_0(p)$ to $G(p)$, while the internal photon propagator is unchanged.
The scalar-photon vertex changes as $p_\mu \to p_\mu-e A_\mu$. Thus
\ba 
\tlscbf \hskip -5pt &=&\hskip -6pt \frac{e^2}{2} \int \frac{d^dp\, d^dq}{(2\pi)^{2d}}\, \frac{1}{(p-q)^2}\left\{ (p+q)^2G(p)G(q)-(ef)^2 \frac{\partial G(p)}{\partial p_\mu}  \frac{\partial G(q)}{\partial q_\mu} \right\}\nn  \\
&=&\hskip -6pt \frac{e^2}{2}  \int \frac{d^dp\, d^dq}{(2\pi)^{2d}}\,  \frac{1}{(p-q)^2}\Bigl\{ [-(p-q)^2+2(p^2+m^2)+2(q^2+m^2)-4 m^2]G(p)G(q)\Bigr. \nn \\
&&\hskip 8cm \Bigl. -(ef)^2 \frac{\partial G}{\partial p_\mu}  \frac{\partial G}{\partial q_\mu} \Bigr\}\nn  \\
&=&\hskip -8pt  -\frac{e^2}{2} \left[\int \frac{d^dp}{(2\pi)^d} G(p)\right]^2\hskip -5pt  -2 e^2 m^2  \int \frac{d^dp\, d^dq}{(2\pi)^{2d}}\, \frac{G(p) G(q)}{(p-q)^2}-2(d-4) e^2(ef)^2  \int \frac{d^dp\, d^dq}{(2\pi)^{2d}}\, \frac{G(p) G(q)}{(p-q)^4}\nn 
\\
&=&\hskip -6pt -\frac{e^2}{2} \Bigl[\quad \olscbf\,\,\,  \Bigr]^2 -2 e^2 m^2  \Bigl[ \quad \tlscbfph\,\,\, \Bigr]
-2(d-4) e^2(ef)^2 \Bigl[ \quad \tlscbfphv\,\,\,  \Bigr]
\label{bkmanip}
\ea
Here we have used the differential equation (\ref{propde}) satisfied by $G(p)$, and in the last term we have integrated by parts to take the momentum derivatives off the scalar propagators and onto the photon propagator. As before, the dot on the propagator means that the propagator is squared.

Notice the similarity between the free-field expression (\ref{freemanip}) and the self-dual background field expression (\ref{bkmanip}). These simple manipulations go some way towards explaining the similarity between the free vacuum diagram relation (\ref{freemanipfull}) and the renormalized background field vacuum diagram relation (\ref{2lscdig}).

We now evaluate, using dimensional regularization, each of the three diagrams appearing on the RHS of (\ref{bkmanip}). 
\ba
\olscbf\,\,\,  = \int \frac{d^dp}{(2\pi)^d} \, G(p)&=&\int_0^\infty dt \frac{e^{-m^2 t}}{\ch^{d/2}(eft)} \int \frac{d^dp}{(2\pi)^d} \, e^{-\frac{p^2}{ef}\tanh(eft)}\nn\\
&=&\frac{(ef)^{d/2-1}}{(4\pi)^{d/2}}\int_0^\infty dy \,\frac{e^{-2\kappa y}}{\sh^{d/2}(y)}\nn \\
&=& -\frac{(m^2)^{1-\epsilon}}{(4\pi)^{2-\epsilon}}\, \frac{\Gamma(1+\epsilon)}{\epsilon(1-\epsilon)}\, \frac{\Gamma(\kappa+1-\frac{\epsilon}{2})}{\Gamma(\kappa+\frac{\epsilon}{2})}\, 
\kappa^{\epsilon -1}
\label{oneloopprop}
\ea
where we recall that $\kappa=m^2/(2ef)$, and we have used the Euler beta function identity
\ba
\int_0^\infty dx\, e^{-b x} \sh^c(x)=\frac{\Gamma(c+1) \Gamma(\frac{b}{2}-\frac{c}{2})}{2^{c+1}\Gamma(\frac{b}{2}+\frac{c}{2}+1)}
\label{integral}
\ea
Note that in the free-field limit, where $\kappa\to\infty$, the background field vacuum bubble diagram in (\ref{oneloopprop}) reduces smoothly to the corresponding free-field vacuum bubble diagram in (\ref{oneloopfreeprop}). Furthermore, for any $\kappa$, as $\epsilon\to 0$ this difference is
\ba
\olscbf \,\,\,  -\quad\olscfr\,\,&=& -\frac{(m^2)^{1-\epsilon}}{(4\pi)^{2-\epsilon}}\frac{\Gamma(1+\epsilon)}{\epsilon(1-\epsilon)}\left(\frac{\Gamma(\kappa+1-\frac{\epsilon}{2})}{\Gamma(\kappa+\frac{\epsilon}{2})}\,\kappa^{\epsilon-1}-1\right)
\nn\\
&=& \frac{m^2}{(4\pi)^2}\left(\psi(\kappa)-\log \kappa +\frac{1}{2\kappa}\right)+O(\epsilon)
\label{xisub}
\ea
which demonstrates how the function $\xi(\kappa)$, defined in (\ref{xi}), arises in dimensional regularization. This relation (\ref{xisub}) was already noted in (\ref{xiint}) with $\epsilon=0$.

To compute the remaining two terms in (\ref{bkmanip}) we use the following identities, which are easily derived in dimensional regularization:
\ba
\int \frac{d^dp\, d^dq}{(2\pi)^{2d}}\, \frac{e^{-p^2 u} e^{-q^2 v}}{(p-q)^2}&=& \frac{2}{(d-2)}\frac{1}{(4\pi)^{d}}\frac{(uv)^{1-d/2}}{(u+v)}
\label{id1}\\
\int \frac{d^dp\, d^dq}{(2\pi)^{2d}} \, \frac{e^{-p^2 u} e^{-q^2 v}}{(p-q)^4}&=&\frac{4}{(d-2)(d-4)}\frac{1}{(4\pi)^{d}}\frac{(uv)^{2-d/2}}{(u+v)^2}
\label{id2}
\ea
Using (\ref{id1}) and the proper-time form (\ref{scprop}) of the scalar propagator in a self-dual background field, we find
\ba
\tlscbfph  &=& \frac{2}{(d-2)} \frac{(ef)^{d-3}}{(4\pi)^d} \int_0^\infty dt \int_0^\infty ds\, e^{-2\kappa(s+t)} \, \frac{[\sh(t)\sh(s)]^{1-d/2}}{\sh(t+s)}\nn\\
&=& \frac{2^{d/2}}{(d-2)} \frac{(ef)^{d-3}}{(4\pi)^d} \int_0^\infty \frac{ y\, dy}{\sh(y)} \, e^{-2\kappa y} \int_0^1 dv\, [\ch( y)-\ch(y v)]^{1-d/2}
\label{pp2}
\ea
where we have made the changes of variables from $s$ and $t$, to $y=(s+t)$, and $v=\left(\frac{s-t}{s+t}\right)$. Similarly, using (\ref{id2}) we find
\ba
\tlscbfphv &=& \frac{4}{(d-2)(d-4)} \frac{(ef)^{d-4}}{(4\pi)^d} \int_0^\infty dt \int_0^\infty ds\, e^{-2\kappa(s+t)} \, \frac{[\sh(t)\sh(s)]^{2-d/2}}{\sh^2(t+s)}\nn\\
&=& \frac{2^{d/2}}{(d-2)(d-4)} \frac{(ef)^{d-4}}{(4\pi)^d} \int_0^\infty \frac{ y\, dy}{\sh^2(y)} \, e^{-2\kappa y} \int_0^1 dv\, [\ch( y)-\ch(y v)]^{2-d/2}
\label{pp4}
\ea
The $v$ integrals in (\ref{pp2}) and (\ref{pp4}) have the same form, and can be done by making the substitution $w=(\ch(y)-\ch(y v))/(\ch(y)-1)$:
\ba
\int_0^1dv \left[\ch(y)-\ch(y v)\right]^\alpha
&=& \frac{(\ch y -1)^{1+\alpha}}{y\,\sh y}\, \int_0^1 dw\, w^\alpha\, (1-w)^{-\frac{1}{2}}\, \left(1-w\, \tanh^2 (\frac{y}{2})\right)^{-\frac{1}{2}}\nn\\
&=&\frac{\Gamma(\frac{1}{2})\Gamma(1+\alpha)}{\Gamma(\frac{3}{2}+\alpha)} \frac{\left(\ch y -1\right)^{1+\alpha}}{y\, \sh y} ~_2F_1\left(1+\alpha,\frac{1}{2},\frac{3}{2}+\alpha ; \tanh^2(\frac{y}{2})\right) \nn\\
&=&
\frac{\Gamma(\frac{1}{2})\Gamma(1+\alpha)}{2^{1+\alpha}\Gamma(\frac{3}{2}+\alpha)} \frac{\left(\sh y \right)^{1+2\alpha}}{y} ~_2F_1\left(\frac{1}{2}+\frac{\alpha}{2},\frac{1}{2}+\frac{\alpha}{2},\frac{3}{2}+\alpha ; -\sh^2 y \right)
\nn
\\
\label{hyper}
\ea
Here we used the hypergeometric identities 2.1.4.(23) and 2.1.5.(25) in \cite{bateman}, but note that there is a typographical error in  2.1.5.(25), as noted at the front of \cite{bateman}.

In (\ref{pp2}) and (\ref{pp4}) the $v$ integrals are of the form (\ref{hyper}), with  $\alpha=1-\frac{d}{2}=\epsilon-1$, and $\alpha=2-\frac{d}{2}=\epsilon$, respectively. Finally, the $y$ integrals can be done by expanding the hypergeometric functions and using the identity (\ref{integral}), leading to
\ba 
\tlscbfph &=&\frac{(m^2)^{1-2\epsilon}}{(4\pi)^{4-2\epsilon}}\, \frac{\Gamma(1+\epsilon) \Gamma(\frac{1}{2})}{\epsilon(1-\epsilon)\Gamma(\frac{\epsilon}{2})^2}\, \kappa^{2\epsilon-1} \sum_{n=0}^\infty \frac{(-1)^n \Gamma(n+\frac{\epsilon}{2})^2 \Gamma(2\epsilon+2n-1)\Gamma(\kappa-n-\epsilon+1)}{n!\, 2^{2n+2\epsilon-1} \Gamma(n+\frac{1}{2}+\epsilon) \Gamma(\kappa+n+\epsilon)}
\nn\\
&=& -\frac{(m^2)^{1-2\epsilon}}{(4\pi)^{4-2\epsilon}}\, \frac{(\Gamma(1+\epsilon))^2}{\epsilon^2 (1-\epsilon)(1-2\epsilon)}\, \frac{\Gamma(\kappa+1-\epsilon)}{\Gamma(\kappa+\epsilon)}\, \kappa^{2\epsilon-1} \left[1+O(\epsilon^3)\right]
\label{pp2epsilon}
\ea
and
\ba 
\tlscbfphv  &=&
-\frac{(m^2)^{-2\epsilon}}{(4\pi)^{4-2\epsilon}}\, \frac{\Gamma(1+\epsilon) \Gamma(\frac{1}{2})}{2\epsilon(1-\epsilon)\Gamma(\frac{1}{2}+\frac{\epsilon}{2})^2}\, \kappa^{2\epsilon} \sum_{n=0}^\infty \frac{(-1)^n \Gamma(n+\frac{1}{2}+\frac{\epsilon}{2})^2 \Gamma(2\epsilon+2n)\Gamma(\kappa-n-\epsilon+\frac{1}{2})}{n!\, 2^{2n+2\epsilon} \Gamma(n+\frac{3}{2}+\epsilon) \Gamma(\kappa+n+\epsilon-\frac{1}{2})}\nn
\\
&=& -\frac{1}{2}\, \frac{(m^2)^{-2\epsilon}}{(4\pi)^{4-2\epsilon}}\, \frac{(\Gamma(1+\epsilon))^2}{\epsilon^2 (1-\epsilon)(1-2\epsilon)(1+2\epsilon)}\, \kappa^{2\epsilon}\, 
\frac{d}{d\kappa}\left( \frac{\Gamma(\kappa+1-\epsilon)}{\Gamma(\kappa+\epsilon)}\right)\,  \left[1+O(\epsilon^2)\right]
\label{pp4epsilon}
\ea
The final expressions in  (\ref{pp2epsilon}) and (\ref{pp4epsilon}) are obtained by noting that 
\ba
~_2F_1\left(\frac{\epsilon}{2},\frac{\epsilon}{2},\frac{1}{2}+\epsilon;-\sh^2(y)\right)&=&1+O(\epsilon^2)
\nn\\
~_2F_1\left(\frac{1}{2}+\frac{\epsilon}{2},\frac{1}{2}+\frac{\epsilon}{2},\frac{3}{2}+\epsilon;-\sh^2(y)\right)&=&\frac{y}{\sh(y)}+O(\epsilon)
\label{hyperexps}
\ea
It is quite remarkable that all the two-loop bare vacuum diagrams for a self-dual background can be expressed so simply in terms of gamma functions. 

\section{Renormalization}

To renormalize the two-loop effective Lagrangian in the dimensional regularization scheme, we first subtract the free vacuum diagram (\ref{freemanip}) from the background field vacuum diagram (\ref{bkmanip}), and then expand in a Laurent series in $\epsilon$, keeping the pole and finite parts. We then identify and subtract the mass and charge renormalization terms from this Laurent series, leaving the renormalized  two-loop effective Lagrangian.

Subtracting the bare vacuum diagram (\ref{freemanip}) from the background field vacuum diagram (\ref{bkmanip}) gives the bare two-loop effective Lagrangian:
\ba
\Bigl[\quad\tlscbf \,\,\,  -\quad \tlscfr \,\,\, \Bigr]_{\rm bare} &=& -\frac{e^2}{2}\Bigl\{\left[ \quad \olscbf\,\,\, \right]^2-\left[\quad \olscfr\,\,\, \right]^2 \Bigr\}-2 e^2 m^2\Bigl\{\quad \tlscbfph \,\,\, - \quad \tlscfrph \,\,\, \Bigr\}  \nn \\     \nn \\
&& -2 (d-4)e^2(ef)^2 \Bigl\{\quad \tlscbfphv \,\,\, \Bigr\} 
\label{bare}
\ea
Using the results  (\ref{oneloopfreeprop}) and (\ref{oneloopprop}), and expanding in $\epsilon$, the first term in (\ref{bare}) is
\ba
-\frac{e^2}{2}\left\{\left[\quad \olscbf\,\,\,  \right]^2-\left[\quad \olscfr\,\,\, \right]^2\right\} &=& -\frac{e^2}{2}\frac{(m^2)^{2-2\epsilon}}{(4\pi)^{4-2\epsilon}}\, \left(\frac{\Gamma(1+\epsilon)}{\epsilon (1-\epsilon)}\right)^2 \left[\left( \frac{\Gamma(\kappa+1-\frac{\epsilon}{2})}{\Gamma(\kappa+\frac{\epsilon}{2})}\, \kappa^{\epsilon-1}\right)^2-1\right]\nn\\
\nn\\
&&\hskip -5cm = -\frac{e^2}{2}\frac{m^4}{(4\pi)^{4}}\left\{\frac{2}{\epsilon}\, \frac{\xi(\kappa)}{\kappa}+\left[  2\frac{\xi^2(\kappa)}{\kappa^2}+4\left(1-\gamma-\log\left(\frac{m^2}{4\pi}\right)\right)\frac{\xi(\kappa)}{\kappa} -\frac{1}{4\kappa^2} \right] +O(\epsilon)\right\}
\label{first}
\ea
Here $\gamma=0.5572...$ is Euler's constant. Using  (\ref{tlfr}) and (\ref{pp2epsilon}), neglecting terms that clearly vanish with $\epsilon$, the second term in (\ref{bare}) is
\ba
-2 e^2 m^2\left\{\quad \tlscbfph\,\,\, - \quad \tlscfrph\,\,\, \right\} &=&2e^2\, \frac{(m^2)^{2-2\epsilon}}{(4\pi)^{4-2\epsilon}}\, \frac{(\Gamma(1+\epsilon))^2}{\epsilon^2 (1-\epsilon)(1-2\epsilon)} \left[ \frac{\Gamma(\kappa+1-\epsilon)}{\Gamma(\kappa+\epsilon)}\, \kappa^{2\epsilon-1}-1\right]\nn\\
\nn\\
&&\hskip -5cm = 2e^2\,\frac{m^4}{(4\pi)^{4}}\left\{\frac{2}{\epsilon}\, \frac{\xi(\kappa)}{\kappa}+\left[  2\frac{\xi^2(\kappa)}{\kappa^2}+4\left(\frac{3}{2}-\gamma-\log\left(\frac{m^2}{4\pi}\right)\right)\frac{\xi(\kappa)}{\kappa} -\frac{1}{2\kappa^2} \right] +O(\epsilon)\right\}
\label{second}
\ea
From (\ref{pp4epsilon}), also neglecting terms that clearly vanish with $\epsilon$, the third term in (\ref{bare}) is
\ba
-2 (d-4)e^2(ef)^2 \left\{\quad \tlscbfphv \,\,\, \right\} &=&-\frac{e^2}{2} \frac{(m^2)^{2-2\epsilon}}{(4\pi)^{4-2\epsilon}}\, \frac{(\Gamma(1+\epsilon))^2}{\epsilon (1-\epsilon)(1-2\epsilon)(1+2\epsilon)} \, \kappa^{2\epsilon-2}\, 
\frac{d}{d\kappa}\left( \frac{\Gamma(\kappa+1-\epsilon)}{\Gamma(\kappa+\epsilon)}\right)
\nn\\
\nn\\
&&\hskip -5cm = -\frac{e^2}{2} \,\frac{m^4}{(4\pi)^{4}}\left\{\frac{1}{\epsilon}\, \frac{1}{\kappa^2}+
\frac{1}{\kappa^2}\left[  2\xi^\prime(\kappa)-1-2\gamma-2\log\left(\frac{m^2}{4\pi}\right) \right] +O(\epsilon)\right\}
\label{third}
\ea
Mass renormalization is effected by subtracting $\delta m^2 \frac{\partial {\mathcal L}^{(1)}}{\partial (m^2)}$ from the bare two-loop Lagrangian,
where $\delta m^2$ is the one-loop mass (squared) shift, computed using dimensional regularization:
\ba
\delta m^2=\left[\quad \olmass \,\,\, \right]_{p^2=-m^2}&=&e^2 \left[\int \frac{d^dq}{(2\pi)^d}\, \frac{(p+q)^2}{(p-q)^2(q^2+m^2)}\right]_{p^2=-m^2}\nn\\
\nn\\
&=& -e^2 \left(\frac{3-2\epsilon}{1-2\epsilon}\right)\frac{(m^2)^{1-\epsilon}}{(4\pi)^{2-\epsilon}}\, \frac{\Gamma(1+\epsilon)}{\epsilon(1-\epsilon)} \nn\\
\nn\\
&=& -\frac{e^2m^2}{(4\pi)^2}\left\{\frac{3}{\epsilon}+\left[7-3\gamma -3\ln\left(\frac{m^2}{4\pi}\right)\right]+O(\epsilon)\right\}
\label{massshift}
\ea
In dimensional regularization, the derivative of the one-loop effective Lagrangian is
\ba
\frac{\partial {\mathcal L}^{(1)}}{\partial (m^2)}&=& \frac{(ef)^{d/2-1}}{(4\pi)^{d/2}}\, \int_0^\infty ds \, e^{-2\kappa s}\left(\frac{1}{\sh^{d/2}(s)}-\frac{1}{s^{d/2}}\right) \nn\\
\nn\\
&=& -\frac{(m^2)^{1-\epsilon}}{(4\pi)^{2-\epsilon}}\, \frac{\Gamma(1+\epsilon)}{\epsilon (1-\epsilon)}\left[  \frac{\Gamma(\kappa+1-\frac{\epsilon}{2})}{\Gamma(\kappa+\frac{\epsilon}{2})}\, \kappa^{\epsilon-1}-1\right]
\label{oneloopderiv}
\ea
Thus the mass renormalization subtraction term is  
\ba
\delta m^2 \frac{\partial {\mathcal L}^{(1)}}{\partial (m^2)}&=& e^2 \left(\frac{3-2\epsilon}{1-2\epsilon}\right)\frac{(m^2)^{2-2\epsilon}}{(4\pi)^{4-2\epsilon}}\, \left(\frac{\Gamma(1+\epsilon)}{\epsilon(1-\epsilon)} \right)^2 \left[  \frac{\Gamma(\kappa+1-\frac{\epsilon}{2})}{\Gamma(\kappa+\frac{\epsilon}{2})}\, \kappa^{\epsilon-1}-1\right]
\nn\\
\nn\\
&& \hskip -3cm =3e^2 \frac{m^4}{(4\pi)^{4}}\left\{\frac{1}{\epsilon}\, \frac{\xi(\kappa)}{\kappa}+\left[  \frac{1}{2}\frac{\xi^2(\kappa)}{\kappa^2}+\left(\frac{10}{3}-2\gamma-2\log\left(\frac{m^2}{4\pi}\right)\right)\frac{\xi(\kappa)}{\kappa} -\frac{1}{8\kappa^2} \right] +O(\epsilon)\right\}
\label{massren}
\ea
Combining the bare Lagrangian (\ref{bare}) with this mass renormalization subtraction, we obtain the mass renormalized two-loop effective Lagrangian:
\ba
\hskip -1cm \Bigl[\quad\tlscbf\,\,\,  -\quad \tlscfr\,\,\, \Bigr]_{\rm mass\,\, ren} &=& 
\Bigl[\quad\tlscbf\,\,\, -\quad \tlscfr\,\,\, \Bigr]_{\rm bare} - \delta m^2 \frac{\partial {\mathcal L}^{(1)}}{\partial (m^2)}\nn\\ \nn\\
&=& e^2 \frac{m^4}{(4\pi)^4}\, \frac{1}{\kappa^2}\left(\frac{3}{2}\xi^2(\kappa)-\xi^\prime(\kappa)\right) +\frac{e^4 f^2}{64 \pi^4}\left(-\frac{1}{2\epsilon}+ \gamma+ \log\left(\frac{m^2}{4\pi}\right)\right)
\label{mrl}
\ea
Notice that in the mass renormalized expression all the $\frac{\xi(\kappa)}{\kappa}$ terms cancel, in both the pole and finite parts,
provided we renormalize {\it on-shell}.
The final term in (\ref{mrl}) is proportional to the bare Maxwell Lagrangian, $f^2$, and so is clearly identified as the two-loop charge renormalization term. The coefficient of the log term in this two-loop charge renormalization term gives us the two-loop $\beta$-function coefficient for scalar QED \cite{ritus,dgs}.
Subtracting the charge renormalization term, the remaining term in (\ref{mrl}) is the final answer for the fully renormalized two-loop effective Lagrangian (\ref{2lsc}) for scalar QED in a constant self-dual background.

\section{Conclusions}

This paper has shown how the fully renormalized two-loop Euler-Heisenberg effective Lagrangian for scalar QED in a self-dual background is naturally expressed in terms of one-loop diagrams. To conclude, we use the explicit computation of the various two-loop diagrams to deduce some diagrammatic rules, which hold to $O(\epsilon^0)$. This leads to a novel diagrammatic perspective on the mass renormalization.  First, recall that the diagrammatic decomposition in (\ref{bkmanip}) was obtained by purely algebraic manipulations, without doing any integrals, and furthermore this decomposition mirrors the corresponding free-field decomposition in (\ref{freemanip}). This leads directly to the diagrammatic representation (\ref{bare}) of the bare two-loop effective Lagrangian. The remaining two-loop diagram combinations appearing in (\ref{bare}) 
are computed in (\ref{second}) and (\ref{third}), and their reduction to one-loop diagrams may be represented diagrammatically as follows:
\ba
\tlscbfph\,\,\, - \,\,\,\, \tlscfrph &=&\hskip-5pt -\frac{1}{m^2} \Bigl\{\left[ \quad \olscbf\,\,\, \right]^2-\left[\quad \olscfr\,\,\, \right]^2 \Bigr\} +\frac{1}{8\pi^2} \Bigl\{ \quad \olscbf\,\, -\,\,\,\, \olscfr\,\, \Bigr\} +\frac{e^2 f^2}{m^2 (4\pi)^4} +O(\epsilon)
\label{rule1}
\ea
\ba
\tlscbfphv &=& -\frac{1}{\epsilon} \frac{1}{(4\pi)^2}
\Bigl\{\quad \mbox{\olscbfv}\,\,\, -\quad\mbox{\olscfrv} \,\, \Bigr\}
+\frac{1}{2\epsilon} \frac{1}{(4\pi)^4}\left(-\frac{1}{\epsilon}+1+2\gamma+2 \log\left(\frac{m^2}{4\pi}\right)\right) +O(\epsilon^0)
\label{rule2}
\ea
Recall that the two-loop bubble in (\ref{rule2}) appears in the two-loop effective Lagrangian (\ref{bare}) multiplied by a factor of $\epsilon$.

These rules mean that the bare two-loop effective Lagrangian in (\ref{bare}) can be written in terms of one-loop diagrams as
\ba
\Bigl[\quad\tlscbf \,\,\,  -\quad \tlscfr \,\,\, \Bigr]_{\rm bare} &=& \frac{3e^2}{2}\Bigl\{\left[ \quad \olscbf\,\,\, \right]^2-\left[\quad \olscfr\,\,\, \right]^2 \Bigr\}- 
e^2\frac{(ef)^2}{4\pi^2}  \Bigl\{\quad \mbox{\olscbfv}\,\,\, -\quad  \mbox{\olscfrv}\,\, \Bigr\}
\nn\\
\nn\\
&&-\frac{e^2m^2}{4\pi^2} \Bigl\{ \quad \olscbf\,\, -\,\,\,\, \olscfr\,\, \Bigr\} 
+\frac{e^4 f^2}{(4\pi)^4} \left( -\frac{2}{\epsilon}+4 \gamma+4 \log\left(\frac{m^2}{4\pi}\right) \right) +O(\epsilon)
\nn\\
\nn\\
&=& 
\frac{3e^2}{2}\left[ \quad \olscbf\,\,\, -\quad \olscfr\,\,\, \right]^2 - 
e^2\frac{(ef)^2}{4\pi^2}  \Bigl\{\quad \mbox{\olscbfv}\,\,\, -\quad  \mbox{\olscfrv}\,\, \Bigr\}
\nn\\
\nn\\
&&\hskip -4cm+\left(3 e^2 \left[\quad\olscfr\,\,\, \right]-\frac{e^2 m^2}{4\pi^2}\right)\Bigl\{ \quad \olscbf\,\, -\,\,\,\, \olscfr\,\, \Bigr\} +\frac{e^4 f^2}{(4\pi)^4} \left( -\frac{2}{\epsilon}+4 \gamma+4 \log\left(\frac{m^2}{4\pi}\right) \right) +O(\epsilon)
\label{baredig}
\ea
where we have "completed the square" to write
\ba 
\left[ \quad \olscbf\,\,\, \right]^2-\left[\quad \olscfr\,\,\, \right]^2 = \left[ \quad \olscbf\,\,\, -\quad \olscfr\,\,\, \right]^2 +2  \left[\quad\olscfr\,\,\, \right] \, \Bigl[ \quad \olscbf\,\, -\,\,\,\, \olscfr\,\,\, \Bigr]
\ea
The first two terms in the last equality in (\ref{baredig}) give the diagrammatic form of the fully renormalized two-loop effective Lagrangian (\ref{2lscdig}). The final term clearly  corresponds to two-loop charge renormalization. The remaining term corresponds to
mass renormalization. To see this, recall that the mass (squared) shift (\ref{massshift}) is
\ba
\delta m^2&=&\left[\quad \olmass \,\,\, \right]_{p^2=-m^2}
\nn\\
&=& e^2 \left(\frac{3-2\epsilon}{1-2\epsilon}\right) \,  \left[\quad\olscfr\,\,\, \right] 
\nn\\
&=& 3 e^2 \left[\quad\olscfr\,\,\, \right]-\frac{e^2 m^2}{4\pi^2}+O(\epsilon)
\label{massshiftdig}
\ea
Thus, the mass renormalization subtraction term is
\ba 
\delta m^2 \frac{\partial {\mathcal L}^{(1)}}{\partial (m^2)}=\left(3 e^2 \left[\quad\olscfr\,\,\, \right]-\frac{e^2 m^2}{4\pi^2}\right)\Bigl\{ \quad \olscbf\,\, -\,\,\,\, \olscfr\,\, \Bigr\} +O(\epsilon).
\label{massrendig}
\ea
We recognize this mass renormalization subtraction term as the third term in (\ref{baredig}), which is obtained after completing the square. This shows that the mass renormalization essentially corresponds to the diagrammatic operation of completing the square. After charge and mass renormalization we are left with the diagrammatic expression (\ref{2lscdig}) for the fully renormalized two-loop effective Lagrangian.

Finally, I re-emphasize the remarkable similarity between the relation (\ref{freemanip}) connecting two-loop and one-loop free vacuum diagrams, and the corresponding relation (\ref{bkmanip}) for vacuum diagrams in a self-dual background field. Furthermore, after mass and charge renormalization we find the same factor of $\frac{3}{2}$ in (\ref{2lscdig}) for the background field case as is found in (\ref{freemanipfull}) for the free case by simple algebraic manipulations. It is surprising that the background field has so little effect, once the two-loop background field vacuum diagram is renormalized on-shell. To make full use of rules like these at higher loop order it would be useful to have a purely  algebraic derivation, for example from generalized  integration-by-parts rules, using the self-dual background field propagators instead of free propagators. Work along these lines is in progress.

{\bf Acknowledgments :} I am grateful to the Rockefeller Foundation for a Bellagio Residency in the summer of 2003, when this work was done. I also thank A. Vainshtein and  C. Schubert for helpful discussions and correspondence, and the DOE for support through the grant DE-FG02-92ER40716.

 \end{document}